\documentclass[prl,amsmath,amssymb,twocolumn,showpacs,floatfix]{revtex4}
\usepackage{graphicx}
\usepackage[english]{babel}
\usepackage{times}
\usepackage{dcolumn}
\usepackage[usenames,dvipsnames]{color}
\usepackage{units}

\begin{document}

\title{Kondo resonance line-shape of magnetic adatoms on
decoupling layers}

\author{Rok \v{Z}itko}

\affiliation{Jo\v{z}ef Stefan Institute, Jamova 39, SI-1000 Ljubljana,
Slovenia,\\
Faculty  of Mathematics and Physics, University of Ljubljana,
Jadranska 19, SI-1000 Ljubljana, Slovenia}

\date{\today}

\begin{abstract}
The zero-bias resonance in the $\mathrm{d}I/\mathrm{d}V$ tunneling
spectrum recorded using a scanning tunneling microscope above a
spin-1/2 magnetic adatom (such as Ti) adsorbed on a decoupling layer
on metal surface can be accurately fitted using the universal spectral
function of the Kondo impurity model both at zero field and at finite
external magnetic field. Excellent agreement is found both for the
asymptotic low-energy part and for the high-energy logarithmic tails
of the Kondo resonance. For finite magnetic field, the nonlinear
fitting procedure consists in repeatedly solving the impurity model
for different Zeeman energies in order to obtain accurate spectral
functions which are compared with the experimental
$\mathrm{d}I/\mathrm{d}V$ curves. The experimental results at zero
field are sufficiently restraining to enable an unprecedented
reliability in the determination of the Kondo temperature, while at
finite fields the results are more ambiguous and two different
interpretations are proposed.
\end{abstract}


\newcommand{\dIdV}{\mathrm{d}I/\mathrm{d}V}
 
\maketitle

When an atom with non-zero spin and/or orbital magnetic moment is
adsorbed on an ultrathin insulating (``decoupling'') layer grown on a
metal substrate, it will behave as a nearly ideal magnetic impurity
with well-defined local moment which is described by the Kondo model
\cite{zener1951,kondo1964,anderson1961,hewson}. The Kondo model
explains the effects of the weak coupling of the impurity moment with
the itinerant electrons in the substrate. If the coupling is
antiferromagnetic, the impurity moment is screened in a complex
many-particle process (the Kondo effect \cite{hewson, wilson1975,
tsvelick1983,andrei1983}), the signature of which is a resonance in
the impurity spectral function which peaks in close vicinity of the
Fermi level \cite{gunnarsson1983b,li1998,madhavan1998,ternes2009}. The
theoretically predicted resonance line-shape is parabolic near its
maximum \cite{nozieres1974, hewson}, has long tails which can be
empirically well fitted by an inverse square root (Doniach-{\v
S}unji{\'c}) function \cite{frota1986,frota1992,bulla2000}, but
which have been shown to really be logarithmic
\cite{logan1998,dickens2001}; the Kondo resonance cannot be well
approximated by a simple Lorentzian resonance \cite{ujsaghy2000,
fu2007kondo,pruser2011}.

Recent tunneling spectroscopy experiments performed at very low
temperatures provide $\dIdV$ spectra of magnetic adatoms with
excellent energy resolution and very low noise
\cite{heinrich2004,hirjibehedin2006,otte2008}. Numerical techniques
for accurately computing impurity spectral functions at finite
frequencies have also reached maturity
\cite{frota1986,sakai1989,costi1994,bulla1998,bulla2000,hofstetter2000,peters2006,weichselbaum2007,
resolution}. We are thus presently able to directly compare
experimental spectra with the theory. This is interesting for several
reason. First, it is a non-trivial test of the adequacy of simplified
quantum impurity models for describing the low-energy behavior of
single magnetic adsorbates. Second, it provides an unbiased approach
for extracting the parameters of the impurity model. Third, it
confirms our understanding of the logarithmic dynamical behavior of
quantum impurities in the intermediate frequency regime (above the
Kondo scale, but below the atomic scale). Our goal in this work is to
fit the experimental tunneling spectra for a Ti adatom on the
Cu$_2$N/Cu(100) surface, Fig.~4 in Ref.~\onlinecite{otte2008}, using
Kondo resonance curves calculated using the numerical renormalization
group (NRG) technique \cite{bulla2008,pruser2011}. The fitting window
is the bias voltage window from \unit[-10]{mV} to \unit[10]{mV}, which
includes not only the Kondo resonance peak but also its long-tailed
flanks.

The experimental Kondo-resonance spectrum for Ti shows a sizeable
asymmetry between positive and negative bias voltages and the
resonance peak is slightly displaced from zero bias. These two
features indicate possible presence of additional potential scattering
in the problem (in addition to the exchange scattering, which is the
crucial element of the Kondo model). It is, however, also possible to
account for the asymmetry using a model which takes into account
quantum interference between a narrow and a broad scattering channel
(similar to the Fano formula used for fitting the spectra of magnetic
impurities adsorbed directly on a metal surface
\cite{madhavan1998,fano1961, ujsaghy2000}, but based on the correct
impurity spectral function rather than on an approximation by a
Lorentzian peak \cite{ujsaghy2000,pruser2011}), while the displacement
of the peak might be explained by some small bias offset $V_0$ for
experimental reasons (such as thermovoltages generated by temperature
differences in the STM head or in the cryostat wiring). As we show in
the following, it is indeed possible to obtain an excellent fit to the
experimental data with the universal Kondo-resonance line-shape of the
symmetric Kondo impurity model; it is not necessary to include any
potential scattering in the impurity model itself.

The fitting is performed with the phenomenological ansatz function
\begin{equation}
\label{eq2}
(\dIdV)(V) = a + b \left[ (1-q^2) \mathrm{Im}\,G(eV)
+ 2 q \mathrm{Re}\,G(eV) \right],
\end{equation}
where the impurity Green's function is chosen to be
\begin{equation}
G(\omega) = G_\mathrm{Kondo}[ (\omega-\omega_0)/\Delta_\mathrm{HWHM} ].
\end{equation}
The ansatz function \eqref{eq2} is chosen so as to reproduce the Fano
formula for the case of a pure potential scatterer. The bias voltage
is expressed in energy units as $\omega=eV$, where $e$ is the
elementary charge; $\omega_0=eV_0$ with the voltage offset $V_0$.
Parameter $a$ accounts for the possible featureless background
conduction, parameters $b$ and $\Delta_\mathrm{HWHM}$ set the height
and width of the resonance, respectively, and $q$ is a Fano-model-like
asymmetry parameter such that $q=0$ describes a symmetric peak, while
$q\neq 0$ leads to an asymmetry. Finally, $G_\mathrm{Kondo}$ is the
universal impurity Green's function for the Kondo model, as computed
using a highly accurate NRG calculation. It is normalized and rescaled
so that in zero magnetic field $A(\omega=0) =
(-1/\pi)\mathrm{Im}\,G_\mathrm{Kondo}(\omega=0)=1$ and
$A(\omega=1)=1/2$. We show in the following that the five-parameter
ansatz in Eq.~\eqref{eq2} is sufficient to obtain nearly perfect
agreement between the experimental results and the theoretical
spectral function at zero magnetic field throughout an extended energy
range including the resonance peak and its tails. We perform nonlinear
fitting using the Levenberg-Marquardt method to obtain the parameter
set $(a,b,\Delta_\mathrm{HWHM},q,\omega_0)$. To quantify the
``goodness of fit'' we calculate squared residuals $\sigma^2$ for all
data points.

\begin{figure}[htbp]
\centering
\includegraphics[clip,width=8cm]{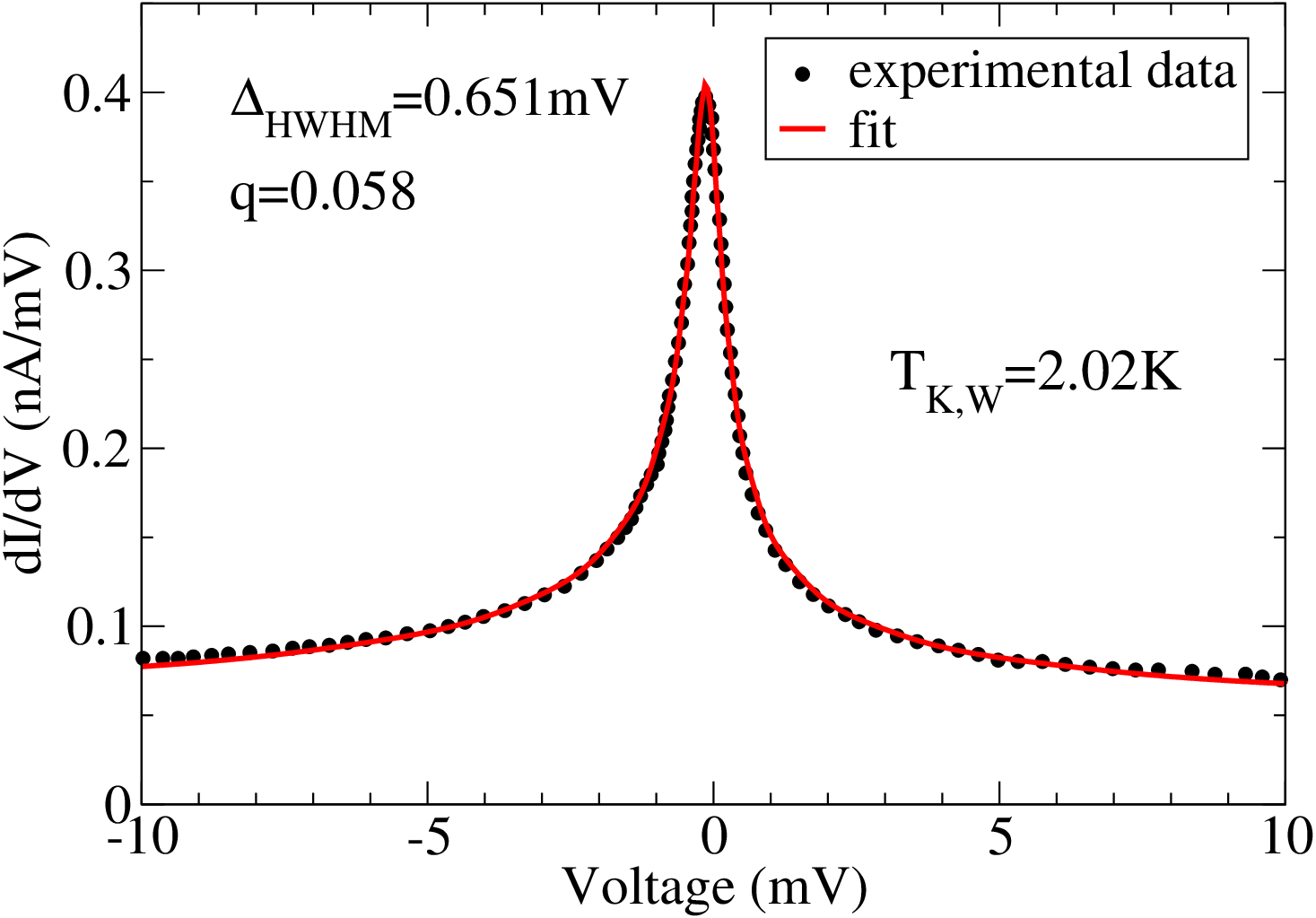}
\caption{(Color online) Theoretical function (light line, red online)
and experimental data (black symbols) for the Kondo resonance in the
$\dIdV$ spectral function for a Ti magnetic impurity atom adsorbed on
the Cu$_2$N layer on Cu(100) surface. The theoretical function is
given by Eq.~\eqref{eq2}, the parameters are explained in the main
text. The experimental results have been extracted from Fig.~4 in
Ref.~\onlinecite{otte2008}.}
\label{fig1}
\end{figure}

We first discuss the Kondo resonance in the absence of the external
magnetic field, $B=0$. The comparison of the theoretical function,
Eq.~\eqref{eq2}, with the experimental data is shown in
Fig.~\ref{fig1} and the extracted parameters are tabulated in
Table~\ref{tab1} (first line). Some comments are in order. First, the
asymmetry parameter $q$ is non-zero, but small. It is not necessary to
add some polynomial background to the model in order to describe the
asymmetry. A linear background hardly affects the goodness of fit
$\sigma^2$. A quadratic background somewhat improves agreement in the
tails, yet -- importantly -- the value of $q$ as extracted using such
an extended fit function changes by only $3\%$. Thus we conclude that
the Fano interference is the correct physical interpretation of the
observed asymmetry. To further support this claim, additional
calculations have been performed for the Kondo model with potential
scattering, the Kondo model with conduction-band density of states
with a finite slope, as well as for the asymmetric single-impurity
Anderson model; the results indicate that it is not possible to
reproduce the observed asymmetry without including the Fano-like
interference in the ansatz function, except for rather extreme (and
thus unphysical) values of the parameters in those models.

The second comment concerns the parameter $\Delta_\mathrm{HWHM}$.
Accurate NRG calculations indicate that the relation between the HWHM
of the spectral function for spin-$1/2$ Kondo model at the
particle-hole symmetric point and the Kondo temperature (as defined by
Wilson) is \cite{resolution}
\begin{equation}
\Delta_\mathrm{HWHM}/T_{K,W} \approx 3.7.
\end{equation}
We thus find $k_B T_{K,W} = \unit[0.176]{meV}$ or
\begin{equation}
T_{K,W} = \unit[2.04]{K}.
\end{equation}
Away from the particle-hole symmetric point the ratio
$\Delta_\mathrm{HWHM}/T_{K,W}$ is different (it depends on the
quasi-particle scattering phase shift) \cite{resolution}. It is also
worth mentioning that there are several different definitions of the
Kondo temperature for the $S=1/2$ Kondo model which are in common use.
The Wilson's definition $T_{K,W}$ is often used in NRG, perturbation
theory, and Bethe Ansatz studies, another definition $T_K^{(0)}$ is
used in local Fermi liquid works, and a third definition $T_{K,H}$ has
been used in the works of Hamann, Nagaoka and Suhl. They are related
through $T_{K,W}=0.4128 T_K^{(0)}$ and $T_{K,H}=2.2 T_{K,W}$. We thus
find
\begin{equation}
\begin{split}
T_K^{(0)} &= \unit[4.94]{K}, \\
T_{K,H} &= \unit[4.49]{K}.
\end{split}
\end{equation}
Some care is needed when comparing works where the Kondo temperature
is defined in non-equivalent ways.

The final third comment concerns the thermal broadening. The
calculation has been performed at $T=0$, while the experiment
\cite{otte2008} is performed at $T=\unit[500]{mK}$, which is smaller
than $T_{K,W}$ only by a factor of four. It is known, however, that
the main effect of finite temperature in the $T < T_K$ range is to
reduce the height on the resonance peak and only slightly increase the
peak width; the logarithmic tails are not affected significantly. This
indicates that the finite-temperature effects are small and that the
reliability of the extracted parameters is not reduced. The NRG
calculation can be performed at finite temperature, but this does not
improve the quality of the fit in the present case. Nevertheless, it
would be interesting to repeat the experiment at even lower
temperature in an attempt to achieve the true asymptotic
zero-temperature limit and to reduce the noise even further.

\begin{table*}[htbp]
\centering
\begin{ruledtabular}
\begin{tabular}{@{}lllllll@{}}
Parameter & $a$ & $b$ & $q$ & $\Delta_\mathrm{HWHM}$ & $\omega_0$ &
$g\mu_B B/T_K$ \\
$B=\unit[0]{T}$    & $0.0239 \pm 0.0008$
& $-0.00472 \pm 0.00002$ 
& $0.058 \pm 0.003$ 
& $\unit[(0.651 \pm 0.008)]{meV}$
& $\unit[(-0.116 \pm 0.003)]{meV}$ 
& 0 (*) \\
$B=\unit[7]{T}$ (A) 
& $0.3112 \pm 0.001$ 
& $-0.00417 \pm 0.00005$ 
& $0.021 \pm 0.007$ 
& $\unit[(0.651 \pm 0.008)]{meV}$\,\, (*)
& $\unit[(-0.16 \pm 0.01)]{meV}$ 
& 5.14 \\
$B=\unit[7]{T}$ (B) 
&$0.3112 \pm 0.0005$ 
&$-0.0052 \pm 0.0002$ 
&$0.029 \pm 0.003$ 
& $\unit[(0.409 \pm 0.002)]{meV}$ 
& $\unit[(-0.133 \pm 0.003)]{meV}$ 
& 7.36  \\
& \end{tabular}
\end{ruledtabular}
\caption{Full set of fit parameters for the Kondo resonance peaks
shown in Figs.~\ref{fig1} and \ref{fig2}. The error estimates
indicated include only the standard deviation from the fitting
procedure. The asterisk indicates a parameter whose value is fixed in
the minimization.} 
\label{tab1}
\end{table*}

\begin{figure}[htbp]
\centering
\includegraphics[clip,width=8cm]{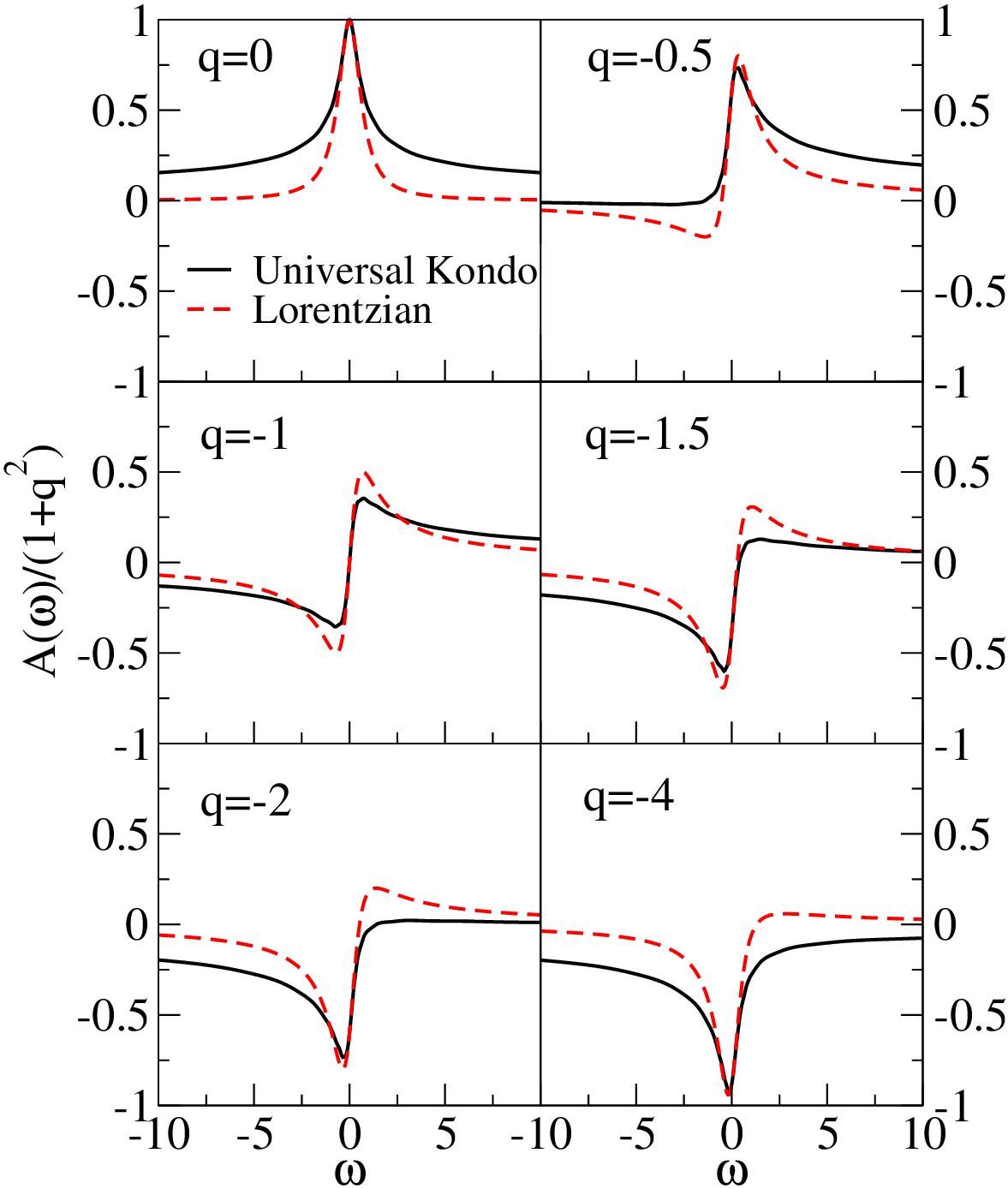}
\caption{(Color online) Comparison of the resonance line-shapes
arising from Fano-like quantum interference in processes where the
resonant channel is described by a Lorentzian (dashed line, red
online) or by the Kondo resonance curve (full line, black online).}
\label{fano}
\end{figure}

For reference, in Fig.~\ref{fano} we plot a number of spectral
line-shapes for Fano-like interference processes for cases where the
narrow resonant scattering channel is described either by a Lorentzian
curve or by a Kondo resonance curve (see also
Refs.~\onlinecite{pruser2011,dargelphd}). The width of the Lorentzian
has been fixed so that in the asymptotic low-energy region both curves
overlap. As a general rule, at the same value of the asymmetry $q$ the
``Fano-Kondo'' line-shapes have longer tails and smoother variation
compared to their ``Fano-Lorentzian'' counterparts. In particular, it
should be noted that the experimental $\dIdV$ curve in Fig.~\ref{fig1}
cannot be well described using a Fano-Lorentzian line-shape.

\begin{figure}[htbp]
\centering
\includegraphics[clip,width=8cm]{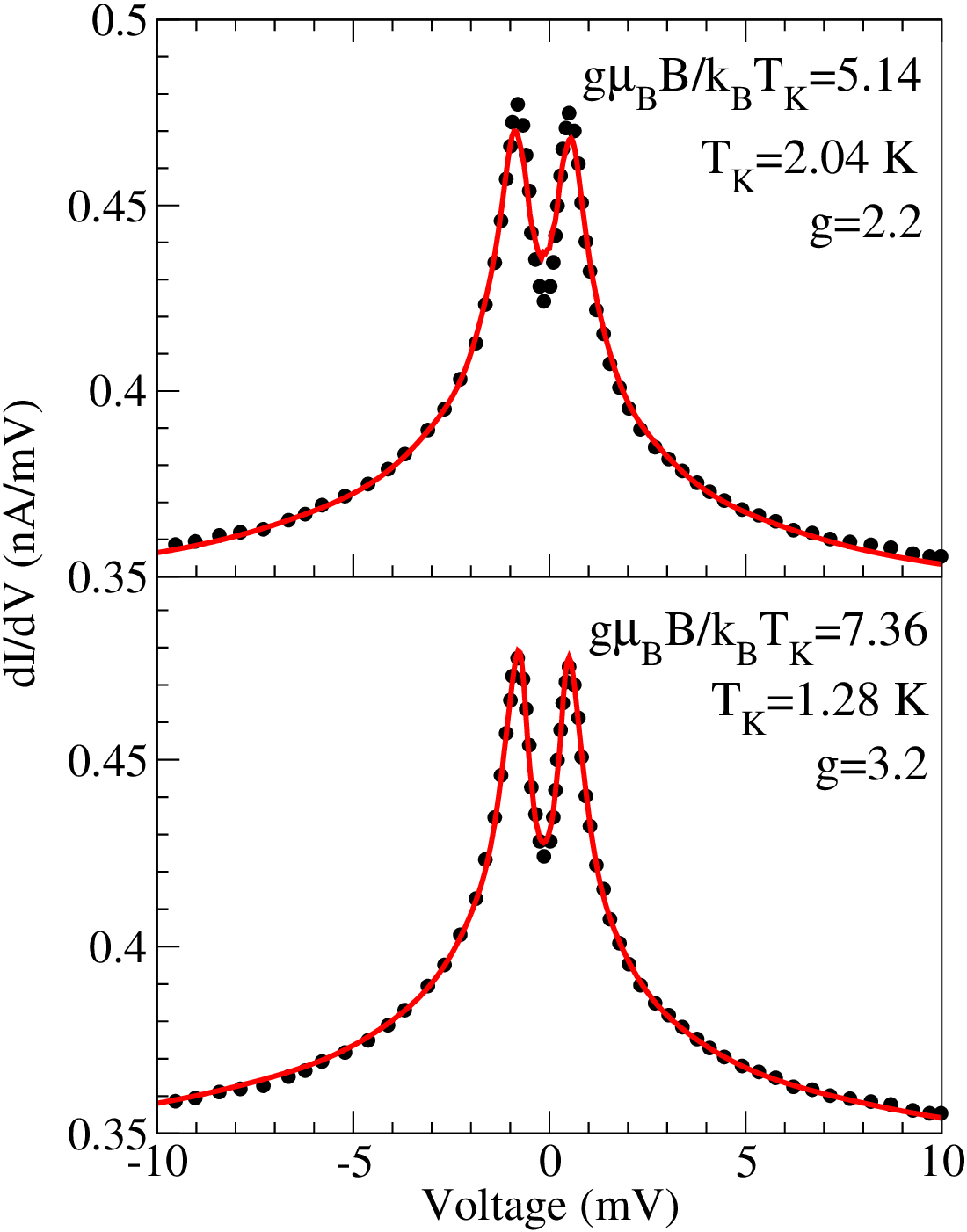}
\caption{(Color online) Theoretical function (light function, red
online) and experimental data (black symbols) for the Kondo resonance
splitting in the external magnetic field $B=\unit[7]{T}$. Upper panel:
fixed parameter $\Delta_\mathrm{HWHM}$. Lower panel: unconstrained
$\Delta_\mathrm{HWHM}$. The magnetic field strength is \unit[7]{B}.
The experimental data have been extracted from Fig.~4 in
Ref.~\onlinecite{otte2008}. } \label{fig2}
\end{figure}

\begin{figure}[htbp]
\centering
\includegraphics[clip,width=8cm]{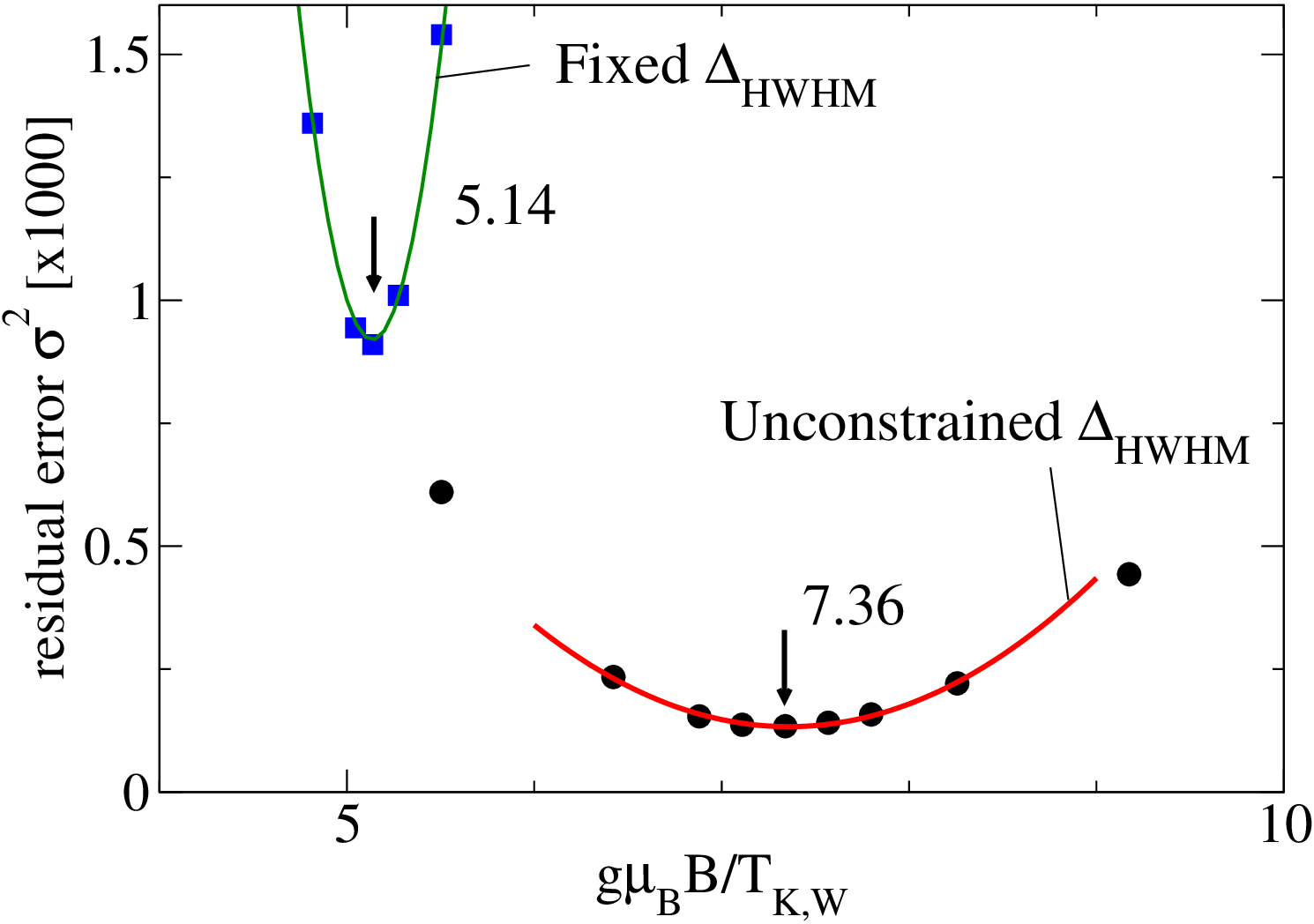}
\caption{(Color online) ``Goodness of fit'' $\sigma^2$ (sum of the
residual squared errors) for the field-splitting of the Kondo
resonance assuming constant or unconstrained parameter
$\Delta_\mathrm{HWHM}$ as a function of the Zeeman energy to
Kondo temperature ratio.}
\label{figerror}
\end{figure}

We now turn to the $\dIdV$ spectra in the presence of the magnetic
field, where the splitting of the Kondo resonance is observed
\cite{costi2000,moore2000,logan2001,hewson2006,quay2007}. We need to
establish the value of a new parameter, the Zeeman energy $b=g\mu_B
B$. We note that while $B$ is known from the experiment, the
$g$-factor is not and needs to be dermined by the fitting procedure
that we now perform. This is a very non-trivial task, since the Kondo
resonance splitting is not linear as a function of the Zeeman energy
\cite{costi2000,moore2000,logan2001,zitko2011}. For each value of the ratio
$b/T_K$ a different universal (split) Kondo resonance curve needs to
be computed \footnote{Spectral functions are provided as Supplemental
Information.}. In minimizing the parameters in the model function,
Eq.~\eqref{eq2}, we first fix $\Delta_\mathrm{HWHM}$: this choice is
equivalent to assuming that the magnetic field does not change the
effective exchange coupling $J_K$ of the impurity spin, which is a
reasonable assumption. The resulting fit is shown in Fig.\ref{fig2}
(top panel), the parameters are tabulated in Table~\ref{tab1} (second
line, A), and the variation of the residual error as a function of
$b/T_K$ is plotted in Fig.~\ref{figerror}. The agreement is fairly
good, but not outstanding: the amplitude of the two resonance peaks is
clearly underestimated. From the known field $B=\unit[7]{K}$ and the
extracted $b/T_K$ ratio, we determine the value of the $g$-factor:
\begin{equation}
g=\mathrm{2.2}.
\end{equation}
This value is in the same range as the $g$-factors for other magnetic
adatoms on the same surface \cite{heinrich2004}.

The fit may be improved by relaxing the constraint that the effective
Kondo exchange coupling $J_K$ does not vary with the magnetic field.
The reason for doing so is the observation that the $S=1/2$ behavior
at low energy scales in Ti adsorbates originates from the freezing out
of the $S=1$ degree of freedom on some higher energy scales; indeed, a
Ti atom in free space has two $3d$ electrons and is in a $S=1$, $L=3$
state \cite{otte2008,griffith}. Upon adsorption on the surface, the
local orbital moment is strongly quenched and the spin moment is also
reduced. The presence of the external magnetic field might affect how
the original impurity degrees of freedom are combined into an
effective $S=1/2$ object and how this effective spin is coupled with
the substrate electron \cite{schrieffer1966}. Since the Kondo
temperature depends exponentially on $J_K$, i.e., $T_K \propto
\exp(-1/\rho J_K)$, where $\rho$ is the density of states in the
substrate, even a small variation of $J_K$ might induce a sizable
change in $T_K$. We thus performed a second fitting calculation where
the parameter $\Delta_\mathrm{HWHM}$ was allowed to change. The result
of this fit is shown in Fig.~\ref{fig2}, bottom panel, and the
parameters are listed in Table~\ref{tab1}, second line (B). The
agreement improves significantly, see also the residual errors plotted
in Fig.~\ref{figerror}. The Kondo temperature determined by this
approach is smaller by 40\% compared to $T_K$ at zero field, and the
$g$-factor is found to be $g=3.2$. While the purported change in $T_K$
in the field is quite large and the value of $g$ is larger than
typical values for other adatoms, it is not clear whether the
improvement of the fit can be solely ascribed to having a further free
parameter in the ansatz function (i.e., overfitting) or if there are
physical grounds for reduced $J_K$ in the effective model. This
ambiguity calls for further systematic experimental studies. 

It has been shown that the zero-bias peaks in $S=1/2$ impurities on
decoupling layers can be accurately described by the universal Kondo
spectral function, confirming the logarithmic frequency-dependence in
the resonance tails. The Kondo temperature has been extracted in a 
reliable way. Due to non-linear behavior of the magnetic-field
splitting of the Kondo resonance, two interpretations of the peak
structure have been proposed which differ in the values of the
exchange coupling $J_K$ and the $g$-factor.

\begin{acknowledgments}
I acknowledge the support of the Slovenian Research Agency (ARRS)
under Grant No. Z1-2058 and Program P1-0044.
\end{acknowledgments}

\bibliography{fit}

\end{document}